\documentclass[preprintnumbers,floatfix,twocolumn,prc]{revtex4}

\usepackage{enumerate}
\usepackage{amsmath,amssymb}  
\usepackage{bm}               
\usepackage{amscd}            
\usepackage{graphicx}         
\usepackage{hhline,multirow}  
\usepackage{dcolumn}          

\newcommand{\beq}{\begin{equation}}
\newcommand{\eeq}{\end{equation}}
\newcommand{\bea}{\begin{eqnarray}}
\newcommand{\beas}{\begin{eqnarray*}}
\newcommand{\beau}[1]{\begin{equation} \label{#1} \begin{array}{rcl}}
\newcommand{\eea}{\end{eqnarray}}
\newcommand{\eeas}{\end{eqnarray*}}
\newcommand{\eeau}{\end{array} \end{equation}}
\newcommand{\bay}{\begin{array}}
\newcommand{\eay}{\end{array}}
\newcommand{\btab}{\begin{tabular}}
\newcommand{\etab}{\end{tabular}}
\newcommand{\bals}{\begin{align*}}
\newcommand{\eals}{\end{align*}}

\newcommand{\ra}{{\rightarrow}}

\newcommand{\vev}[1]{\langle #1 \rangle}

\newcommand{\nbar}{{\overline n}}

\begin{document}


\preprint{JLAB-THY-08-867}

\title{Target mass corrections for spin-dependent structure
	functions \\ in collinear factorization}

\author{Alberto~Accardi$^{a,b}$ and W. Melnitchouk$^{b}$}
\affiliation{
$^a$Hampton University, Hampton, VA 23668, USA \\
$^b$Jefferson Lab, Newport News, VA 23606, USA \\
\\
}

\begin{abstract}
We derive target mass corrections (TMC) for the spin-dependent nucleon
structure function $g_1$ and polarization asymmetry $A_1$ in collinear
factorization at leading twist.
The TMCs are found to be significant for $g_1$ at large $x_B$, even at
relatively high $Q^2$ values, but largely cancel in $A_1$.
A comparison of TMCs obtained from collinear factorization and from
the operator product expansion shows that at low $Q^2$ the corrections 
drive the proton $A_1$ in opposite directions.
\end{abstract}



\maketitle

\section{Introduction}

Understanding the transition from the perturbative to the nonperturbative 
regimes of Quantum Chromodynamics (QCD) remains one of the most 
challenging problems in nuclear and hadron physics.
Recent progress in describing this transition has focused on quark-hadron 
duality, which relates observables computed from quark and gluon degrees 
of freedom to those parametrized in terms of hadronic variables.
A classic example of this is the phenomenon of Bloom-Gilman duality
\cite{BG}, in which the inclusive structure functions in the region
dominated by low-lying nucleon resonances follow deep inelastic
structure functions describing high energy data, to which the resonance
structure functions average \cite{MEK}.

In QCD the observation of this duality can be formulated within the
operator product expansion (OPE), in which moments of structure
functions are expanded in inverse powers of $Q^2$, the four-momentum
squared of the exchanged photon.
The leading term is given by matrix elements of twist-two
local operators, and is associated with single parton
scattering, while the ${\cal O}(1/Q^2)$ and higher terms are related
to higher twist nonperturbative multi-parton correlations.
(The twist of a local operator in the OPE is defined as its mass 
dimension minus its spin.)
The magnitude of the higher twist contributions then determines
the degree to which duality holds \cite{DGP}.

In order to reliably extract information on the duality-violating
higher twist contributions to structure functions, it is vital to
remove from the data kinematical corrections associated with nonzero
values of $Q^2/\nu^2 = 4 x_B^2 M^2/Q^2$, where $\nu$ is the energy
transfer, $M$ is the nucleon mass, and $x_B = Q^2/2M\nu$ is the
Bjorken scaling variable.
While formally related to twist-two operators \cite{Nachtmann},
these ``target mass corrections'' (TMCs) are suppressed by powers
of $M^2/Q^2$, hence TMCs are sometimes inaccurately referred to as
``kinematical higher twists'', and can obscure information on genuine
higher twist terms.

The importance of TMCs has been highlighted recently by 
high-precision data from Jefferson Lab on both spin-averaged and 
spin-dependent structure functions \cite{JLab} taken at moderate
$Q^2$ values, $Q^2 \sim 1-5$~GeV$^2$, and at large $x_B$, where TMCs
are most significant.
Furthermore, with high-intensity neutrino-nucleus scattering
experiments planned in similar kinematics \cite{MINERvA}, TMCs
for weak interactions also need to be understood.

Target mass corrections for spin-averaged nucleon structure functions
were first considered by Georgi \& Politzer within the OPE \cite{DGP,GP}, 
and later extended to the full set of electroweak structure functions
\cite{Blumlein,KretzerOPE}.
For spin-dependent scattering, these were evaluated within the same OPE 
formalism in Refs.~\cite{Matsuda:1979ad,Piccione:1997zh}, extended to
the full set by Bl\"umlein \& Tkabladze \cite{Blumlein}, and computed 
by Detmold \cite{Detmold} for the deuteron.

One of the limitations of the OPE formulation of TMCs is the so-called 
``threshold problem'', in which the target mass corrected structure 
functions remain nonzero at $x_B \geq 1$.
This arises from the failure to consistently incorporate the elastic
threshold in moments of structure functions at finite $Q^2$, resulting 
in nonuniformity of the $Q^2 \to \infty$ and $n \to \infty$
limits, where $n$ is the rank of the moment.
After performing an inverse Mellin transform on the moments, the
extracted structure functions consequently acquire incorrect support
at large $x_B$ \cite{Tung}.
A number of attempts have been made to redress the threshold problem
by considering various prescriptions to tame the unphysical behavior
as $x_B \to 1$ \cite{Tung,Steffens,Kulagin}.
These approaches are not unique, however, and sometimes introduce
additional complications (see Ref.~\cite{Schienbein} for a review).

An alternative approach, which avoids the threshold ambiguities from
the outset, involves formulating TMCs directly in momentum space
\cite{Accardi} using the collinear factorization (CF) formalism
\cite{EFP,Collins}.
This method was heuristically applied by Aivazis, Olness \& Tung
\cite{Aivazis} and by Kretzer \& Reno \cite{KretzerCF} to
spin-averaged structure functions.
More recently Accardi \& Qiu \cite{Accardi} applied this formalism
to deep inelastic structure functions at large $x_B$, carefully taking
into account the elastic threshold and thereby solving the threshold
problem. However, in the handbag approximation, without introducing a
suitable jet function accounting for the invariant mass of the final
hadronic state \cite{Accardi}, leading order structure functions can
still be nonzero at $x_B=1$.

In this Letter we use the CF framework to derive target mass corrections 
to the leading twist $g_1$ and $g_2$ structure functions and the $A_1$ 
polarization asymmetry.
In Sec.~\ref{sec:CF} we outline the main steps in the derivation;
a more detailed account will be presented elsewhere \cite{Accardi_full}.
In Sec.~\ref{sec:numest} we compare and contrast the predictions for 
the TMCs in CF with those using the standard OPE formulation.
We find that the TMCs using the CF method are generally larger for
the $g_1$ structure function than in the OPE.
However, since the TMCs are qualitatively similar for $g_1$ and $F_1$,
the effects largely cancel in the $A_1$ asymmetry, although the residual
effects can still be up to 20\% at large $x_B$, and for $A_1$ even 
differ in sign for the CF and OPE approaches.
Finally, in Sec.~\ref{sec:conclusion} we summarize our findings and 
preview future work.

\section{TMC in collinear factorization}
\label{sec:CF}

The computation of TMCs in collinear factorization makes use of the
factorization theorem relating the hadronic tensor $W^{\mu\nu}$ for
$\gamma^*N$ scattering to the partonic tensor $w^{\mu\nu}_f$ for the
scattering of a virtual photon from a parton of flavor $f$.
The target mass corrected structure functions are then obtained by
suitable projections of the hadronic tensor without neglecting the
target mass $M$ relative to $Q^2$ at any stage.

The hadronic tensor for spin-dependent inclusive scattering of
leptons from nucleons is given by
\begin{align}
\begin{split}
& W^{\mu\nu}(p,q) 
  = \frac{1}{p\cdot q} \varepsilon^{\mu\nu\rho\sigma} q_\rho \\
& \quad \times \Big[ S_\sigma g_1(x_B,Q^2) 
    + \Big( S_\sigma - \frac{S\cdot q}{p\cdot q} p_\sigma
	\Big) g_2(x_B,Q^2)
  \Big] \ ,
\label{eq:Wmunu}
\end{split}
\end{align}
where $p$ and $q$ are the target nucleon and virtual photon 
four-momenta, respectively, and $S$ is the nucleon spin vector,
with $S^2 = - M^2$ and $S \cdot p = 0$.
We work in collinear frames, defined such that $p$ and $q$ do not have
transverse momentum.
This allows us to decompose $p$, $q$ and the parton four-momentum $k$
in terms of light-cone vectors $n^\mu$ and $\nbar^\mu$ as \cite{EFP}
\begin{align}
\begin{split}
  p^\mu &= p^+ \nbar^\mu 
         + \frac{M^2}{2 p^+} n^\mu\ ,		\\
  q^\mu &= - \xi p^+ \nbar^\mu 
         + \frac{Q^2}{2\xi p^+} n^\mu\ ,	\\
  k^\mu &= x p^+ \nbar^\mu 
         + \frac{k^2 + k_T^2}{2 x p^+} n^\mu 
         + k_T^{\,\mu} \ .
 \label{eq:kinematics} 
\end{split}
\end{align}
where $n^2 = \nbar^2 = 0$ and $n \cdot \nbar=1$.
The transverse parton momentum vector $k_T^{\,\mu}$
satisfies $k_T \cdot n = k_T \cdot \nbar = 0$.

The nucleon plus-momentum $p^+ = (p_0 + p_3)/\sqrt{2}$ can be
interpreted as a parameter for boosts along the $z$-axis, connecting
the target rest frame to the hadron infinite-momentum frame.
In terms of the plus-components of the momenta, the parton fractional
light-cone momentum is defined as $x = k^+ / p^+$, while the virtual
photon fractional momentum
\begin{align}
  \xi = - \frac{q^+}{p^+} 
      = \frac{2 x_B}{1 + \sqrt{1+\gamma^2}}
\end{align}
coincides with the Nachtmann scaling variable \cite{Nachtmann},
with $\gamma^2 = 4x_B^2 M^2/Q^2$.
In the Bjorken limit ($Q^2 \ra \infty$ at fixed $x_B$), $\xi \to x_B$
and we recover the standard kinematics in the $M \approx 0$ 
approximation.

The nucleon polarization vector $S$ can be decomposed into longitudinal
($S_L$) and transverse ($S_T$) components,
\begin{align}
  S^\mu = \sigma \lambda\ S_L^\mu + S_T^\mu \ ,
 \label{eq:Sdecomp}
\end{align}
where $S_L^2 = -M^2$, $S_L\cdot p = S_T\cdot p=0$, and the nucleon 
helicity $\lambda = \pm 1$ indicates polarization parallel or 
antiparallel to the nucleon direction of motion.
The degree of longitudinal polarization is given by
$\sigma = \sqrt{1 + S_T^2/M^2}$, with the limits $\sigma_{\rm min} = 0$
and $\sigma_{\rm max} = 1$ describing nucleons with purely transverse
($S_T^2 = -M^2$) or longitudinal ($S_T^2 = 0$) polarization,
respectively.

Collinear factorization for the hadronic tensor can be obtained by
expanding the parton momentum $k$ around its on-shell
($k^2 \to m_f^2=0$) and collinear ($k_T \to 0$) component,
\begin{align}
  k^\mu\ \to\ \widetilde k^\mu = x p^+ \nbar^\mu\ .
  \label{eq:ktilde}
\end{align}
In terms of the on-shell parton momentum $\widetilde k$ we can then
define the collinear invariant
\begin{align}
  x_f = \frac{-q^2}{2\widetilde k\cdot q} = \frac{\xi}{x} \ ,
 \label{eq:tildexf}
\end{align}
where the second equality hold for massless quarks, to which we
restrict this analysis.

The spin vector of a collinear parton, $s^\mu$, is defined analogously
such that $s^2 = 0$ and $s \cdot \widetilde k = 0$.
For a massless spin-1/2 quark, as well as for a massless spin-1 gluon,
the spin vector can be written as
\begin{align}
  s^\mu = \lambda_f \lambda\ \widetilde k^\mu \ ,  
\end{align}
where the parton helicity $\lambda_f = \pm 1$ corresponds to a parton 
with spin parallel or antiparallel to the proton longitudinal spin.

According to the QCD factorization theorem \cite{Collins}
the hadronic tensor can be factorized as
\begin{equation}
W^{\mu\nu}(p,q,S)
= \sum_{f,\lambda_f} 
    \int_\xi^{\xi/x_B} \frac{dx}{x} \, 
    w_f^{\mu\nu}(\widetilde k,q,s) \,
    \varphi_f^{\lambda_f}(x,Q^2)\ ,
 \label{eq:pQCDfactWSA}
\end{equation}
where $w_f^{\mu\nu}$ is the partonic tensor for scattering
from a parton of flavor $f$.
(Note that the vector $s$ in the argument of $w_f^{\mu\nu}$ 
depends on $\lambda_f$.)
The upper limit of integration in Eq.\eqref{eq:pQCDfactWSA}, {\em viz.},
$x_{\rm max} = \xi/x_B$, guarantees that structure functions vanish
for $x_B > 1$ \cite{Accardi}, in contrast to 
Refs.~\cite{Aivazis,KretzerCF} where $x_{\rm max} = 1$.
The neglected terms of order ${\cal O}(k^\mu - \widetilde k^\mu)$ in the
collinear expansion are suppressed by powers of $\Lambda^2/Q^2$, with
$\Lambda$ some hadronic scale, and contribute to the restoration of gauge
invariance in higher-twist diagrams \cite{Qiu:1988dn}.
The factorized expression \eqref{eq:pQCDfactWSA} is obtained with the
additional approximation of neglecting the intrinsic parton $k_T$
and parton off-shellness in the kinematics of the handbag diagram.
The $(x_B,Q^2)$ region where this approximation is valid has been
estimated in Ref.~\cite{Accardi}.  A detailed account of non-zero
$k_T$ requires going beyond the collinear factorization formalism
used in the present analysis.

Polarized scattering is described by the antisymmetric part of
the tensor, which can be decomposed in terms of the partonic
$g_{1,f}$ and $g_{2,f}$ structure functions,
\begin{align}
\begin{split}
& w_f^{\mu\nu}(\widetilde k,q) 
  = \frac{1}{\widetilde k\cdot q} \varepsilon^{\mu\nu\rho\sigma} q_\rho \\
& \quad \times \Big[ s_\sigma g_{1,f}(x_f,Q^2) 
    + \Big( s_\sigma
	  - \frac{s\cdot q}{\widetilde k\cdot q} \widetilde k_\sigma 
      \Big) g_{2,f}(x_f,Q^2)
  \Big] \ .
\label{eq:wfmunu}
\end{split}
\end{align}
For ease of notation, in the following we will omit the dependence on
$Q^2$ of the structure functions and parton distributions functions.
The function $\varphi_f^{\lambda_f}$ in Eq.~(\ref{eq:pQCDfactWSA})
is the parton distribution function for a parton of flavor
$f$ and helicity $\lambda_f$ inside a nucleon.
In the light-cone gauge, and at leading order in $\alpha_s$,
this is defined as
\begin{align}
\begin{split}
\varphi_f^{\lambda_f}(x)
& = \int \frac{dz^-}{2 \pi} e^{-ix p^+ z^-} \\
& \times \vev{p,S|\overline\psi_f(z^-n)\,
	   \frac12 (1 + \lambda_f \gamma_5) \frac{\gamma^+}{2}\,
           \psi_f(0)|p,S}\ ,
\label{eq:quarkhelPDFatLO}
\end{split}
\end{align}
where $\psi_f$ is the quark Dirac field. 
For polarized scattering the spin-dependent quark distribution
function $\Delta\varphi_f$ is then given by
\begin{equation}
\Delta\varphi_f(x)
= \frac{1}{\sigma} \big[ \varphi^+_f(x) - \varphi^-_f(x) \big]\ .
\end{equation}

Note that the factorized expression in Eq.~\eqref{eq:pQCDfactWSA}
is suitable for discussing the contribution of helicity parton 
distributions to the hadronic tensor at leading order in the expansion
of parton correlators in powers of $1/p^+$ \cite{Bacchetta,Jaffe,JiOsb}.  
Extension to transversity distributions, or inclusion of higher order
corrections in $1/p^+$ (corresponding to ``dynamical twist'' $\ge3$ in
the language of Refs.~\cite{Bacchetta,Jaffe}), require a generalization 
of Eq.~\eqref{eq:pQCDfactWSA}.
Even at ${\cal O}(1/p^+)$ the TMCs can become nontrivial \cite{EFP},
and we will discuss these higher order corrections elsewhere.

Using suitable projection operators, structure functions can be 
projected from the hadronic and partonic tensors in 
Eqs.~(\ref{eq:Wmunu}) and (\ref{eq:wfmunu}), and using
Eq.~(\ref{eq:pQCDfactWSA}) one finds 
\begin{align}
g_1(x_B) &= {1 \over 1+\gamma^2} \sum_f 
	    \int_\xi^{\xi/x_B} \!\frac{dx}{x}\,
	    g_{1,f}\left({\xi\over x}\right)\
	    \Delta\varphi_f(x)\ ,
\label{eq:pTMC_CF1} \\
g_2(x_B) &= - g_1(x_B)
\label{eq:pTMC_CF2}
\end{align}
for the target mass corrected structure functions at ${\cal O}(1)$
in $1/p^+$.
At leading order in $\alpha_s$ the partonic structure function
$g_{1,f}$ is proportional to $\delta(x-\xi)$, in which case the
target mass corrected nucleon $g_1$ structure function is given by
\begin{equation}
\label{eq:g1TMCLO}
g_1(x_B) = {1 \over 1 + \gamma^2} g_1^{(0)}(\xi)\ ,
\end{equation}
where $g_1^{(0)}$ is the structure function in the massless target limit, 
$M^2/Q^2 \ra 0$.
Note that Eq.~(\ref{eq:g1TMCLO}) is strictly valid only at leading order.
At higher orders the massless limit $g_1$ structure function generalizes 
to
\begin{equation}
g_1^{(0)}(x_B)
= \sum_f \int_{x_B}^1 \frac{dx}{x}\,
  g_{1,f}\left({x_B\over x}\right)\,
  \Delta\varphi_f\Big(\frac{x_B}{x}\Big)\ ,
\label{eq:TMC_CF0}
\end{equation}
with $g_2^{(0)}(x_B) = - g_1^{(0)}(x_B)$.
Clearly, in general one has $g_1(x_B) \ne g_1^{(0)}(\xi)$ because
of the $1/(1+\gamma^2)$ factor in Eq.~(\ref{eq:pTMC_CF1}), and the 
upper limits of integration ({\em i.e.}, $\xi/x_B$ versus 1).

In actual polarized deep-inelastic scattering experiments one typically 
measures not the structure functions directly, but the virtual photon
polarization asymmetries $A_1$ and $A_2$, defined as ratios of 
spin-dependent to spin-averaged structure functions,
\begin{eqnarray}
  A_1(x_B) &=& \frac{g_1(x_B) - \gamma^2 g_2(x_B)}{F_1(x_B)}\ ,
\label{eq:A1} \\
  A_2(x_B) &=& \gamma\frac{g_1(x_B)+g_2(x_B)}{F_1(x_B)}\ ,
\label{eq:A2} 
\end{eqnarray}
Using the results in Eqs.~(\ref{eq:pTMC_CF1})--(\ref{eq:pTMC_CF2}) one
can write the asymmetries in collinear factorization 
as
\begin{eqnarray}
A_1(x_B) &=& (1+\gamma^2) \frac{g_1(x_B)}{F_1(x_B)}\ ,  \label{eq:A1CF}\\
A_2(x_B) &=& 0\ ,					\label{eq:A2CF}
\end{eqnarray}
where $F_1$ is the spin-averaged structure function, which in
collinear factorization is given by \cite{Accardi}
\begin{equation}
F_1(x_B) = \sum_f \left( F_{1,f} \otimes \varphi_f \right)(\xi)\ ,
\end{equation}
using a shorthand notation $\otimes$ for the integral over $x$
as in Eq.~(\ref{eq:pTMC_CF1}).
The function $\varphi_f$ is defined as the sum of the
helicity distributions in Eq.~(\ref{eq:quarkhelPDFatLO}),
$\varphi_f(x) = \varphi_f^+(x) + \varphi_f^-(x)$.

The polarization asymmetry in collinear factorization can then be 
written at ${\cal O}(1)$ in the $1/p^+$ expansion as
\begin{equation}
A_1(x_B)
= { \sum_f \left( g_{1,f} \otimes \Delta\varphi_f \right)(\xi)
    \over 
    \sum_f \left( F_{1,f} \otimes \varphi_f \right)(\xi) }\ .
\end{equation}
Note that the $(1+\gamma^2)$ prefactor is absent if the asymmetry
is written in terms of the parton distributions directly.
In the $M^2/Q^2 \to 0$ limit the asymmetries are given by
\begin{equation}
  A_1^{(0)}(x_B) = {g_1^{(0)}(x_B) \over F_1^{(0)}(x_B)}\ ,\ \ \ \
  A_2^{(0)}(x_B) = 0 \ .
  \label{eq:A10} 
\end{equation}

The massless $A_1^{(0)}$ asymmetry is also directly related to the 
lepton asymmetry $A_\parallel$ for scattering leptons with longitudinal 
polarization aligned and anti aligned with the nucleon polarization,
\begin{align}
\label{eq:A0Apar}
  A_1^{(0)}(x_B) & = \frac{A_\parallel(x_B)}{D}\ ,
\end{align}
where $D$ is a depolarization factor of the virtual photon \cite{Lampe}.
A commonly used approximation in experimental data analysis relates the 
longitudinal lepton asymmetry with the ratio of the $g_1$ and $F_1$
structure functions,
\begin{equation}
  (1+\gamma^2)\frac{g_1}{F_1} \approx \frac{A_\parallel}{D}\ .
 \label{eq:A1approx}
\end{equation}
From Eqs.~(\ref{eq:A1CF}) and (\ref{eq:A0Apar}) this is equivalent to
assuming that 
\begin{align}
  A_1 \approx A_1^{(0)} \ .  
\end{align}
In the next section we shall test the validity of this approximation
numerically, and compare the results of the collinear factorization
with the target mass corrections obtained from the OPE.

\section{Comparison with the OPE}
\label{sec:numest}

\begin{figure*}[htb]
\centering
  \includegraphics
    [width=\linewidth,clip=true]
    {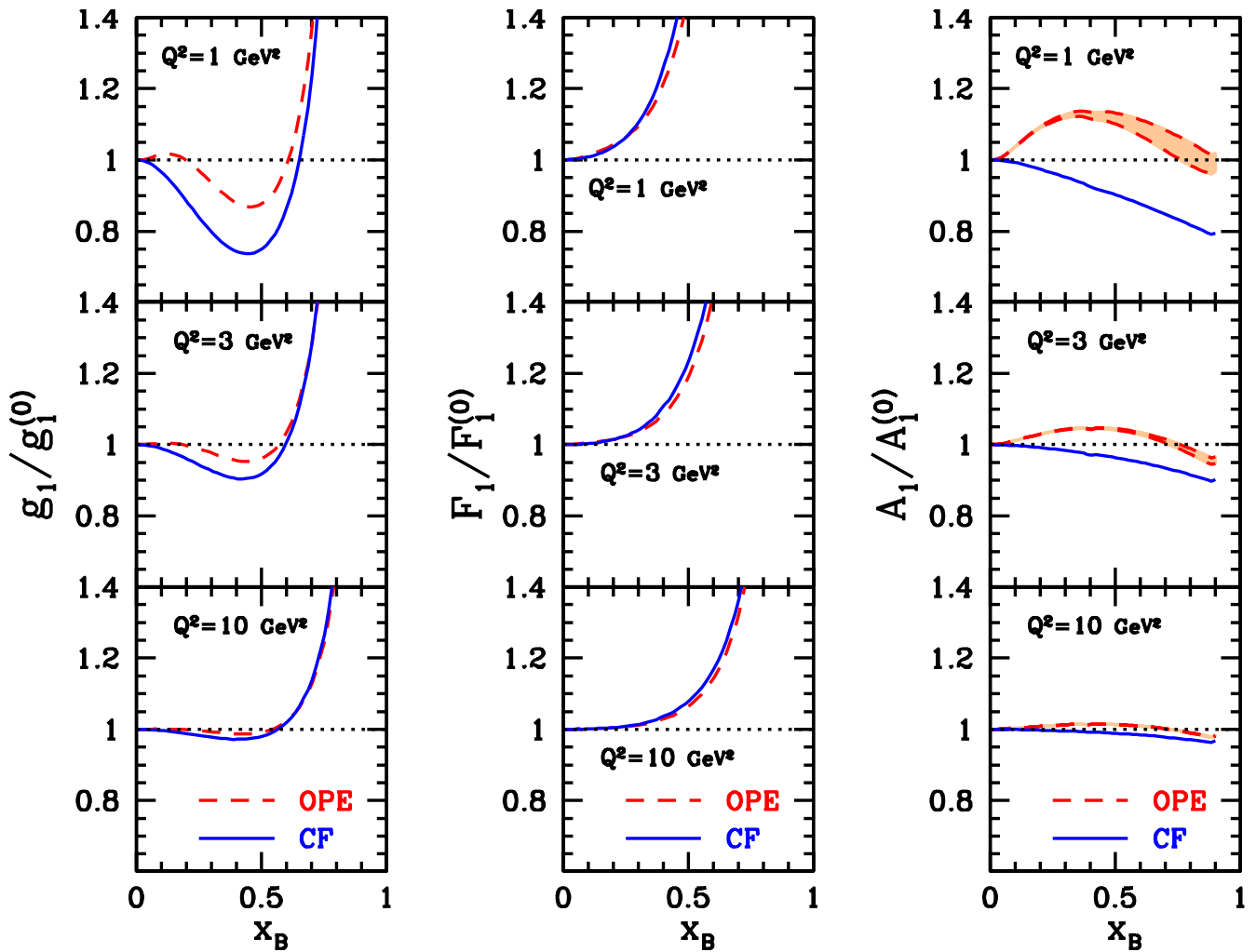}
  \caption{Ratio of target mass corrected to massless proton $g_1$
    (left panel) and $F_1$ (center panel) structure functions and $A_1$
    polarization asymmetry (right panel) in collinear factorization
    (solid) and in the OPE (dashed), for $Q^2=1$, 3 and 10~GeV$^2$.
    For $A_1$ the shaded band for the OPE result indicates the
    effect of using the Wandzura-Wilczek relation, Eq.~\eqref{eq:WW} 
    (lower bound), or the identity $g_1+g_2=0$, Eq.~\eqref{eq:pTMC_CF2} 
    (upper bound).} 
  \label{fig:g1A1_fixedQ2}
\end{figure*}

\begin{figure*}[tbh]
\centering
  \includegraphics
    [width=0.75\linewidth,trim=0 250 0 0,clip=true]
    {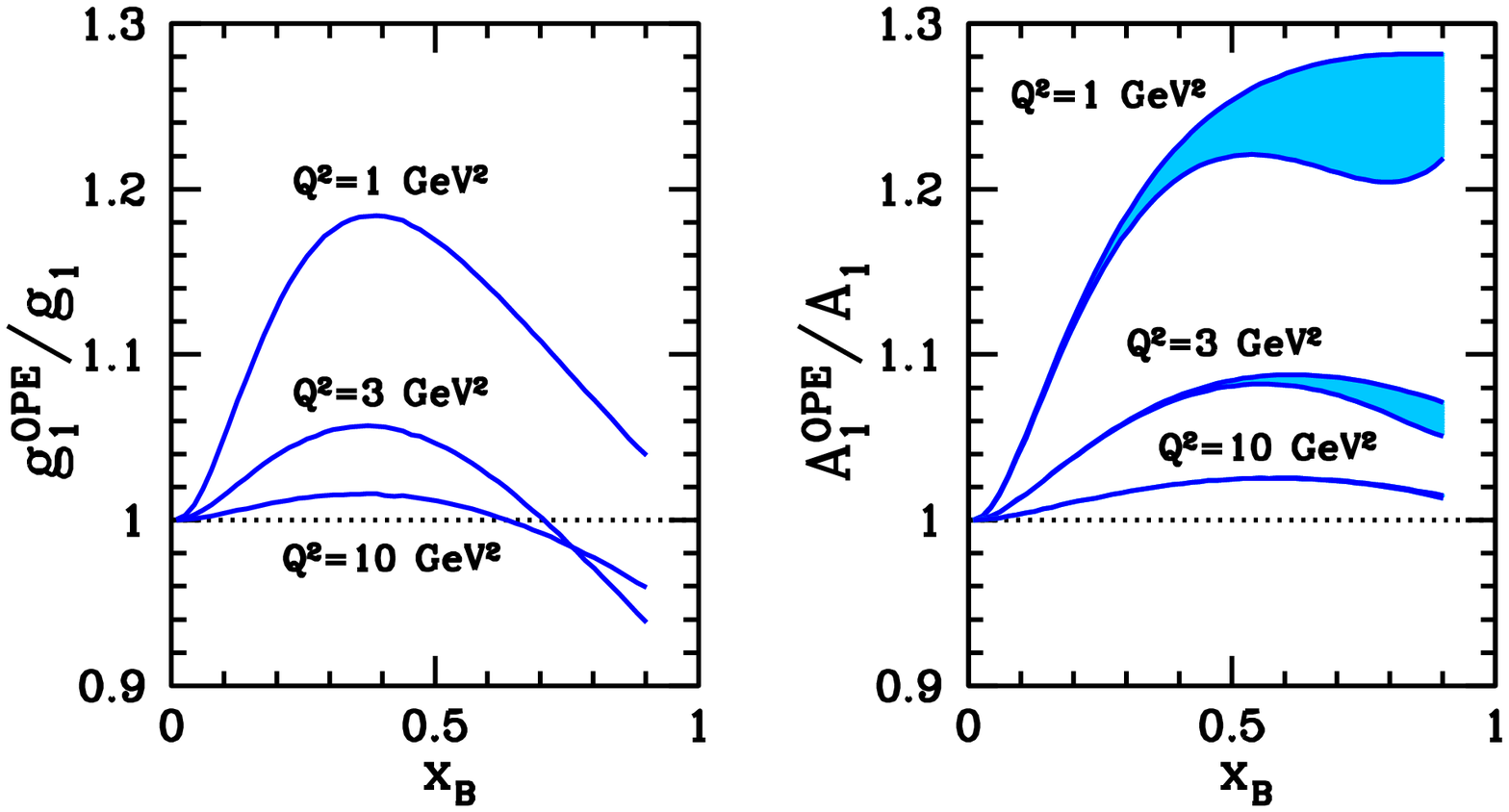}
  \caption{Ratio of the $g_1$ structure functions (left panel) and $A_1$ 
    polarization asymmetries (right panel) computed with target mass 
    corrections in the OPE and CF formalisms at $Q^2=1$ (largest ratios), 
    5 and 10~GeV$^2$ (smallest ratios).
    For $A_1$ in the OPE is as in Fig.~\ref{fig:g1A1_fixedQ2}.}
  \label{fig:OPEPM_fixedQ2}
\end{figure*}

A common prescription for evaluating target mass corrections uses the 
operator product expansion to compute moments of structure functions
at leading twist, including the trace terms which introduce the 
kinematical $M^2/Q^2$ corrections, and extracts the TMC structure 
functions through an inverse Mellin transform 
\cite{DGP,GP,Piccione:1997zh,Blumlein,Schienbein}.
The resulting target mass corrected $g_1$ and $g_2$ structure functions
can be written as \cite{Blumlein}:
\begin{align}
  g_1^\text{OPE}(x_B) 
    & = \frac{1}{(1+\gamma^2)^{3/2}} \frac{x_B}{\xi} g_1^{(0)}(\xi) 
			\nonumber\\
+ \frac{\gamma^2}{(1+\gamma^2)^2} &
  \int_\xi^1\! \frac{dv}{v} 
    \Big[ \frac{x_B+\xi}{\xi} + \frac{\gamma^2-2}{2\sqrt{1+\gamma^2}} 
    \log\big( \frac{v}{\xi}\big) \Big] g_1^{(0)}(v) \ ,
			\label{eq:g1OPE} \\
  g_2^\text{OPE}(x_B)
    & = - g_1^\text{OPE}(x_B)\
      +\  \int_{x_B}^1 \frac{dy}{y} g_1^\text{OPE}(y) \ .
			\label{eq:WW}
\end{align}
The expression for $g_2^{\rm OPE}$ in Eq.~\eqref{eq:WW} is known as
the Wandzura-Wilczek relation \cite{WW}, and was shown in 
Ref.~\cite{Blumlein} to survive target mass corrections.
This expression differs from Eq.~\eqref{eq:pTMC_CF2}, obtained at
${\cal O}(1)$ in the $1/p^+$ expansion in collinear factorization,
by the presence of the integral term.
In collinear factorization such a term emerges at ${\cal O}(1/p^+)$,
and Eq.~\eqref{eq:WW} holds if one neglects quark-gluon-quark correlators
\cite{Mulders,Tangerman} and matrix elements related to the Wilson line
in the expansion of the quark-quark correlators \cite{Goeke}.

The prefactors for $g_1^{(0)}$ in the first term of Eq.~\eqref{eq:g1OPE} 
differ from those in the corresponding collinear factorization 
expression, Eqs.~\eqref{eq:pTMC_CF1} and \eqref{eq:g1TMCLO}.
The factor $(1+\gamma^2)^{-1}$ can be traced back to the tensor
decomposition of $W^{\mu\nu}$ and has the same origin as the factor
appearing in Eq.~\eqref{eq:pTMC_CF1}, while the remaining 
$(1+\gamma^2)^{-1/2}x_B/\xi$ factor arises from the OPE treatment of TMCs. 
Substituting Eq.~\eqref{eq:WW} in Eq.~\eqref{eq:A1} one obtains for the
$A_1$ asymmetry in the OPE:
\begin{align}
\begin{split}
  & A_1^\text{OPE}(x_B) \\
  & \quad = \frac{(1+\gamma^2)}{F_1^\text{OPE}(x_B)} 
    \bigg[ g_1^\text{OPE}(x_B) 
    - \gamma^2 \int_{x_B}^1 \frac{dy}{y} g_1^\text{OPE}(y) \bigg]\ ,
      \label{eq:A1OPE}
  \end{split}
\end{align}
which again differs from Eq.~\eqref{eq:A1CF} in the integral term.
One should also note that the Wandzura-Wilczek relation \eqref{eq:WW}
is not a direct consequence of the OPE \cite{Jaffe,Anselmino},
which leaves open the possibility of $\delta$-function contributions
at $x_B=0$ to the right-hand-side of Eq.~(\ref{eq:WW}).
To explore the phenomenological consequences of contributions to $A_1$
from subleading powers of $1/p^+$, we consider both definitions in the
numerical evaluation of $A_1$.

In Fig.~\ref{fig:g1A1_fixedQ2} we compare the results of a leading order
evaluation of target mass corrected versus uncorrected proton $g_1$ 
(left panel) and $F_1$ (center panel) structure functions, and
polarization asymmetries $A_1$ (right panel) for several $Q^2$ values,
using the leading order GRSV2000 (standard scenario) polarized parton
distributions for $g_1$ \cite{Gluck00} and the GRV98LO unpolarized
distributions for $F_1$ \cite{Gluck98}. 
For both the CF and OPE corrections, the $g_1$ ratio dips below unity
at intermediate $x_B$, $0.2 \lesssim x_B \lesssim 0.5$, before rising
dramatically at larger $x_B$.
The magnitude of the dip and the steepness of the rise for are
naturally greater at lower $Q^2$.
However, while the size of the TMCs at $x_B \lesssim 0.5$ is
$\lesssim 2$--3\% for $Q^2 > 10$~GeV$^2$, at larger $x_B$ the
corrections remain significant even at much larger $Q^2$.
For these reasons, the commonly adopted cut $Q^2 > 1$~GeV$^2$ for
polarized parton distribution function analysis requires inclusion
of TMCs for extracting precise PDFs.

For the $A_1$ polarization asymmetry the TMC effects largely cancel
in the ratio because the TMCs in the $F_1$ structure function are
similar to those in $(1+\gamma^2)g_1$
\cite{GP,KretzerOPE,Accardi,KretzerCF},  
Nevertheless, the residual effects can still be up to 20\% at
$Q^2 = 1$~GeV$^2$, decreasing to $\sim 2$--3\% at $Q^2 = 10$~GeV$^2$.
This provides a quantitative test of the validity of the commonly
used approximation in Eq.~(\ref{eq:A1approx}) for the longitudinal
asymmetry $A_\parallel$ in terms of $A_1$.
Interestingly, the TMC effects drive $A_1$ in opposite directions
for $x_B \lesssim 0.7$, with $A_1$ increasing relative to $A_1^{(0)}$ 
in the OPE approach but decreasing in the CF formulation.
This is due to the fact that for most $x_B$ values $g_1^\text{OPE} >
g_1^\text{CF}$, while $F_1^\text{OPE} < F_1^\text{CF}$.
Such ordering arises mainly from the different prefactors for $g_1$
in Eq.~\eqref{eq:pTMC_CF1} and in the first term of Eq.~\ref{eq:g1OPE}
(for the analogous formulas for $F_1$ see Ref.~\cite{Accardi}).

The effect of using the Wandzura-Wilczek relation in the computation
of $A_1$ instead of the CF result $g_1 = -g_2$, indicated by the shaded 
band in Fig.~\ref{fig:g1A1_fixedQ2}, has $\lesssim 5\%$ effect in general, 
and is negligible for $Q^2 \gtrsim 3$~GeV$^2$. 
The contribution of the Wandzura-Wilczek term is small compared with
the differences between the two TMC schemes considered.

The differences between the two TMC implementations can be seen more
dramatically in Fig.~\ref{fig:OPEPM_fixedQ2}, where the ratios of
OPE and CF target mass corrected $g_1$ (left panel) and $A_1$
(right panel) are presented.
At $Q^2 = 1$~GeV$^2$ the $g_1$ structure function corrected using
the OPE prescription can be up to $\sim 20\%$ larger than that using
the CF approach at $x_B \sim 0.4$, with the difference decreasing at
larger $x_B$.
The differences diminish with increasing $Q^2$, so that by
$Q^2 = 10$~GeV$^2$ the methods give essentially the same results
at the 2\% level for all $x_B \lesssim 0.8$.

The polarization asymmetry is similarly found to be up to $\sim 20-30\%$
larger within the OPE approach at $x_B \geq 0.7$ for $Q^2 = 1$~GeV$^2$,
depending on the prescription used for $g_2$, but again decreasing to
$\lesssim 2\%$ for $Q^2 = 10$~GeV$^2$.
In all cases the TMCs are larger for $A_1$ in the OPE than in the CF
approach.
These results clearly highlight the need for a careful treatment of
TMCs in the low-$Q^2$ and large-$x_B$ kinematics.

\section{Conclusion}
\label{sec:conclusion}

In this study we have derived for the first time the target mass 
corrections to the spin-dependent nucleon $g_1$ and $g_2$ structure 
functions, as well as to the polarization asymmetry $A_1$, in the 
framework of collinear factorization.
In the CF framework the threshold problem affecting the OPE framework
is naturally avoided by directly implementing four-momentum conservation
in the handbag diagram, rendering the structure functions zero for
$x_B > 1$.  A further advantage of this formalism is that it can be
readily extended to processes such as semi-inclusive DIS, where the OPE
is not available, and indeed to any other hard scattering process.
Additional corrections to structure functions at large $x_B$, such as
from jet mass corrections or threshold resummation, can also be
naturally incorporated together with TMCs.

The numerical results for the target mass corrections to the leading-order
$g_1$ structure function in CF are found to be qualitatively similar to
those obtained from the OPE, but up to 20\% larger at $Q^2 = 1$~GeV$^2$.
The corrections become smaller at larger $Q^2$, with differences between
the CF and OPE results $\lesssim 2$--3\% for $Q^2 = 10$~GeV$^2$.
Nevertheless, the TMCs remain significant at $x_B > 0.7$ even for
$Q^2 > 10$~GeV$^2$, and need to be taken into account when analyzing
large-$x_B$ data. 
The numerical difference between the two schemes is likely to increase
in a next-to-leading order computation, where the convolution over $x$
in Eq.~\eqref{eq:pTMC_CF1} is performed only up to $\xi/x_B$ instead
of 1 \cite{Accardi}.

Since the TMCs are qualitatively similar for the $g_1$ and $F_1$ 
structure functions, they largely cancel in the $A_1$ asymmetry,
although the sign of the correction is opposite in the CF and OPE 
approaches over most of the range of $x_B$.
The CF target mass effects in $A_1$ can be as large as 20\% at
$Q^2 = 1$~GeV$^2$ for $x_B \sim 0.8$--0.9, again decreasing to less
than a few percent by $Q^2 = 10$~GeV$^2$.
The commonly used approximation relating $A_1$ directly to the
longitudinal asymmetry $A_\parallel$ will therefore break down at
low $Q^2$, so that accurate determination of polarized structure 
functions will require measurement of both $A_\parallel$ and the 
transverse asymmetry $A_\perp$ \cite{Lampe,Anselmino}.

In the future, this analysis can be extended in several directions.
Firstly, while the CF formalism avoids unphysical regions in dealing
with the threshold problem, the corrected structure functions remain
nonzero at $x_B=1$.
To tame this behavior one can follow the approach of Ref.~\cite{Accardi}
by introducing jet mass corrections, which render the TMC structure
functions zero in the limit $x_B \to 1$.
Furthermore, while we have restricted ourselves to massless quarks,
the generalization to heavy flavors can be accommodated within 
the collinear factorization framework.
In addition, future quantitative analysis of large-$x_B$ and low-$Q^2$ 
data will require TMCs to be computed for structure functions at 
subleading powers in $1/p^+$, which will be necessary for a more
complete treatment of $g_2$, for instance.
Finally, work is currently in progress \cite{SIDIS} to extend the
collinear factorization formalism to semi-inclusive deep inelastic
scattering, where the target and hadron mass effects are yet to be
evaluated, and it will be interesting to address the case of transverse 
momentum dependent parton distributions \cite{Bacchetta}, which will
enable the role of the parton intrinsic transverse momentum to be
quantified. \\


\begin{acknowledgments}
We are grateful to A.~Bacchetta and M.~Schlegel for many informative
discussions, and to A.~Metz for helpful correspondence.
This work was supported by the DOE contract No. DE-AC05-06OR23177,
under which Jefferson Science Associates, LLC operates Jefferson Lab,
and NSF award No. 0653508.
\end{acknowledgments}


\end{document}